\documentclass[aps,prd,preprintnumbers,superscriptaddress,nofootinbib]{revtex4}
\usepackage{amsmath, amssymb, amsthm}
\usepackage{mathtools}
\usepackage{mathrsfs}
\usepackage{bm}
\usepackage{slashed}   
\usepackage{graphicx}
\usepackage{multirow}
\usepackage{tikz}
\usepackage[caption=false]{subfig}
\usepackage{relsize}	
\usepackage{array}
\usepackage{float}
\usepackage{color}
\usepackage{xcolor}
\usepackage{soul}
\usepackage{siunitx}
\usepackage{hyperref}
\hypersetup{
}
\allowdisplaybreaks

\def\be{\begin{eqnarray}}
\def\ee{\end{eqnarray}}

\def\dd{{\mathrm{d}}}
\def\imag{{\mathrm{i}}}

\mathchardef\-="2D

\newcommand{\half}[1][1] {\mathsmaller{\frac{#1}{2}}}

\usepackage{hyperref}

\hypersetup{
colorlinks=true,linkcolor=red,anchorcolor=blue,citecolor=blue, filecolor=blue,urlcolor=red,bookmarksnumbered=true, pdfview=FitB
}

\begin{document}

\title{Pion to photon transition form factors with basis light-front quantization}

\author{Chandan Mondal}
\email{mondal@impcas.ac.cn} 
\affiliation{Institute of Modern Physics, Chinese Academy of Sciences, Lanzhou 730000, China}
\affiliation{School of Nuclear Science and Technology, University of Chinese Academy of Sciences, Beijing 100049, China}

\author{Sreeraj Nair}
\email{sreeraj@impcas.ac.cn} 
\affiliation{Institute of Modern Physics, Chinese Academy of Sciences, Lanzhou 730000, China}
\affiliation{School of Nuclear Science and Technology, University of Chinese Academy of Sciences, Beijing 100049, China}

\author{Shaoyang~Jia}\email{syjia@anl.gov}
\affiliation{Department of Physics and Astronomy, Iowa State University,
Ames, IA 50011, U.S.A.}
\affiliation{Physics Division, Argonne National Laboratory, Argonne, IL 60439, U.S.A.}

\author{Xingbo Zhao}\email{xbzhao@impcas.ac.cn}
\affiliation{Institute of Modern Physics, Chinese Academy of Sciences, Lanzhou 730000, China}
\affiliation{School of Nuclear Science and Technology, University of Chinese Academy of Sciences, Beijing 100049, China}

\author{James~P.~Vary}\email{jvary@iastate.edu}
\affiliation{Department of Physics and Astronomy, Iowa State University,
Ames, IA 50011, U.S.A.}

\collaboration{BLFQ Collaboration}

\date{\today}
\begin{abstract}
We obtain the distribution amplitude (DA) of the pion from its light-front wave functions in the basis light-front quantization framework. This light-front wave function of the pion is given by the lowest eigenvector of a light-front effective Hamiltonian consisting a three-dimensional confinement potential and the color-singlet Nambu--Jona-Lasinion interaction both between the constituent quark and antiquark. The quantum chromodynamics (QCD) evolution of the DA is subsequently given by the perturbative Efremov-Radyushkin-Brodsky-Lepage evolution equation. Based on this DA, we then evaluate the singly and doubly virtual transition form factors in the space-like region for $\pi^0\rightarrow \gamma^*\gamma$ and $\pi^0\rightarrow \gamma^*\gamma^*$ processes using the hard-scattering formalism.
Our prediction for the pion-photon transition form factor agrees well with data reported by the Belle Collaboration. However, in the large $Q^2$ region it deviates from the rapid growth reported by the BaBar Collaboration. Meanwhile, our result on the $\pi^0\rightarrow \gamma^*\gamma^*$ transition form factor is also consistent with  other theoretical approaches and agrees with the scaling behavior predicted by perturbative QCD.
\end{abstract}


\maketitle
\section{Introduction}
The parton distribution amplitudes (DAs) that play essential roles in describing the various hard exclusive processes of quantum chromodynamics (QCD) bound states~\cite{Lepage:1980fj,Chernyak:1983ej,Brodsky:1989pv} via the factorization theorem \cite{Collins:1996fb} are among the most basic structure functions. The DAs are therefore complementary to the parton distribution functions (PDFs) associated with inclusive processes~\cite{Sutton:1991ay,Wijesooriya:2005ir,Gluck:1999xe,Aicher:2010cb}. Since the DAs are  longitudinal projections of the hadronic wave functions obtained by integrating out the transverse momenta of the partons \cite{Lepage:1980fj,Chernyak:1983ej,Efremov:1979qk}, they
carry information on QCD bound states at the amplitude level.
Specifically, the lowest moments of the DAs for a quark
and an antiquark inside a meson are closely related to decay constants and transition form factors \cite{Polyakov:2009je,Noguera:2010fe,Terschlusen:2010gtc,Petrov:1998kg}.

The meson-photon transitions with one or two virtual photons are the simplest decay processes in QCD, reflecting the structure of the meson. The associated transition form factors (TFFs) are  crucial for determining important observables, for example  the hadronic light-by-light contribution to the Standard Model prediction of the muon anomalous magnetic moment~\cite{Jegerlehner:2009ry,Nyffeler:2016gnb,Gerardin:2016cqj,Aoyama:2020ynm,Hoferichter:2018dmo,Gerardin:2019vio,Hoferichter:2018kwz}, and the rates of rare pseudoscalar ($P$)  decays: $P\to l\bar{l}~(l\equiv e,\,\mu)$~\cite{Husek:2015wta,Hoferichter:2021lct}. The single-virtual TFFs in the space-like regions up to a large momentum transfer $(Q^2\sim 40$ GeV$^2$) have been measured experimentally by several collaborations~\cite{Behrend:1990sr,Gronberg:1997fj,Denig:2014mma,Uehara:2012ag,Aubert:2009mc,BABAR:2011ad}. 
The results from the BaBar Collaboration~\cite{Aubert:2009mc} demonstrate a rapid growth of $Q^2F_{\pi\gamma}(Q^2)$ in the large $Q^2$ region, ($F_{P\gamma}(Q^2)$ being the meson to photon TFF).  However, the  measurement by the Belle Collaboration~\cite{Uehara:2012ag} shows that $Q^2F_{\pi\gamma}(Q^2)\to$ constant for $Q^2>15$ GeV$^2$. The data from the Belle Collaboration are consistent with the prominent features of perturbative QCD (pQCD)~\cite{Lepage:1980fj,Braaten:1982yp}, where TFFs are expected to follow the asymptotic behavior of, $Q^2F_{P\gamma}(Q^2)\to{\rm constant}$ as  $Q^2\to\infty$. Unlike their results showing growth in $Q^2F_{\pi\gamma}(Q^2)$, the BaBar data~\cite{BABAR:2011ad} for the $Q^2F_{(\eta,\,\eta')\gamma}(Q^2)$ appear consistent with the pQCD predication. The inconsistency between the Belle and the BaBar data for $Q^2 F_{\pi\gamma}(Q^2)$ and the disparities in the behaviors of $Q^2 F_{\pi\gamma}(Q^2)$  and $Q^2F_{(\eta,\eta^{\prime})\gamma}(Q^2)$ at the high $Q^2$ regime as reported by the BaBar Collaboration
have led to various theoretical studies
~\cite{Polyakov:2009je,Mikhailov:2009kf,Radyushkin:2009zg,
Dorokhov:2013xpa,Wu:2010zc, Kroll:2010bf,Agaev:2012tm,
Roberts:2010rn,Brodsky:2011xx,Stefanis:2012yw,Lucha:2011if,
deMelo:2013zza,Agaev:2014wna,Choi:2017zxn,Choi:2007yu,Choi:1997iq,Stefanis:2020rnd,Gerardin:2016cqj,Ahmady:2018muv,Zhong:2021epq,Gao:2021iqq,Ding:2018xwy,Raya:2016yuj}. 


While the measurements of the single-virtual TFFs having been carried out by several collaborations, information is less available on the double-virtual TFFs. Recently,  the BaBar Collaboration~\cite{BaBar:2018zpn} has measured for the first time the double-virtual $\gamma^*(q_1)\gamma^*(q_2)\to\eta'$ TFF, $F_{\eta'\gamma^*}(Q^1_1,Q^2_2)$,  in the space-like region where $Q^2_{1(2)}=-q^2_{1(2)}>0$. Meanwhile, the meson ($M$) TFFs for the doubly virtual $M\to\gamma^*\gamma^*$ transitions have been studied within the Dyson-Schwinger and Bethe-Salpeter framework~\cite{Weil:2017knt}, the chiral perturbation theory~\cite{Bickert:2020kbn}, a light-front quark model~\cite{Choi:2019wqx}, the anti-de Sitter (AdS)/QCD~\cite{Brodsky:2011yv,Stoffers:2011xe}, and the lattice QCD~\cite{Gerardin:2016cqj,Lin:2013im,Feng:2012ck,Dudek:2006ut,Chen:2016yau}.

Our theoretical framework for  meson structures is based on basis light front quantization (BLFQ), which provides an approach for solving
relativistic many-body bound state structure in quantum field theories~\cite{Vary:2009gt,Li:2021jqb,Zhao:2014xaa,Wiecki:2014ola,Li:2015zda,Li:2017mlw,Jia:2018ary,Lan:2019vui,Lan:2019rba,Lekha,Tang:2018myz,Tang:2019gvn,Xu:2019xhk,Xu:2021wwj,Lan:2019img,Qian:2020utg,Lan:2021wok}. In this work, we evaluate the DA of the pion using the light-front wave functions (LFWFs) based on BLFQ~\cite{Vary:2009gt}, within the valence Fock sector of the pion. We then evaluate the singly and doubly virtual TFFs in the space-like region for $\pi^0\rightarrow \gamma^*\gamma$ and $\pi^0\rightarrow \gamma^*\gamma^*$ transitions following the hard-scattering formalism. The effective Hamiltonian includes a three dimensional confining potential consisting of the light-front holography in the transverse direction \cite{Brodsky:2014yha}, a longitudinal confinement~\cite{Li:2015zda,Li:2017mlw}, and the color-singlet Nambu--Jona-Lasinio (NJL) interactions~\cite{Klimt:1989pm,Shigetani:1993dx}. The color-singlet NJL interactions  account for the dynamical chiral symmetry breaking of QCD. The nonperturbative solutions for the LFWFs given by the recent BLFQ study of light mesons~\cite{Jia:2018ary} have been applied successfully to predict the electromagnetic form factors and associated charge radii, PDFs, structure functions and generalized parton distributions of the pion~\cite{Jia:2018ary, Lan:2019rba,Lan:2019vui,Lekha}. Here, we extend our investigations of the pion to compute its  singly and doubly virtual photon TFFs.

\section{BLFQ-NJL model for the light mesons}\label{sc:BLFQ_NJL}
The structures of the bound states are embedded in the LFWFs obtainable as the solutions of the eigenvalue equation of the Hamiltonian
$H_{\mathrm{eff}}\vert \Psi\rangle=M^2\vert \Psi\rangle$, 
where $H_{\mathrm{eff}}$ is the effective Hamiltonian, with $M^2$ being the mass squared eigenvalue of the state $\vert \Psi\rangle$. Within the current modeling of the meson structure through the framework of BLFQ~\cite{Vary:2009gt}, we consider an effective light-front Hamiltonian and solve for its mass eigenvalues and  eigenstates at the model scale  suitable for low-resolution probes. In the valence Fock sector, the effective Hamiltonian for the light mesons from Ref.~\cite{Jia:2018ary} is given by
\begin{align}
H_\mathrm{eff} = \frac{\vec k^2_\perp + m_q^2}{x} + \frac{\vec k^2_\perp+m_{\bar q}^2}{1-x}
+ \kappa^4 \vec \zeta_\perp^2  - \frac{\kappa^4}{(m_q+m_{\bar q})^2} \partial_x\big( x(1-x) \partial_x \big)+H^{\rm eff}_{\rm NJL}\,,\label{eqn:Heff}
\end{align}
where $m_q$ ($m_{\bar q}$) is the mass of the quark (antiquark), and $\vec{k}_\perp$ is the relative transverse momentum. The parameter $\kappa$ represents the strength of the confinement. The transverse confinement is adopted from the light-front holography, where the holographic variable is defined as~$\vec \zeta_{\perp} \equiv \sqrt{x(1-x)} \vec r_\perp$~\cite{Brodsky:2014yha}. The variable $\vec{r}_\perp$ is conjugated to $\vec{k}_\perp$ and measures the transverse separation between the quark and antiquark. The $x$-derivative in the longitudinal confining potential~\cite{Li:2015zda} is defined as $\partial_x f(x, \vec\zeta_\perp) = \partial f(x, \vec \zeta_\perp)/\partial x|_{\vec\zeta_\perp}$. The effective Hamiltonian also includes $H_{\mathrm{NJL}}^{\mathrm{eff}}$ that corresponds to the color-singlet NJL interaction to account for the chiral dynamics~\cite{Klimt:1989pm}.
For the positively-charged pion, the NJL interaction is given by~\cite{Jia:2018ary}
\begin{align}
H_{\mathrm{NJL},\pi}^{\mathrm{eff}} & =G_\pi\, \big\{\overline{u}_{\mathrm{u}s1'}(p_1')u_{\mathrm{u}s1}(p_1)\,\overline{v}_{\mathrm{d}s2}(p_2)v_{\mathrm{d}s2'}(p_2')+ \overline{u}_{\mathrm{u}s1'}(p_1')\gamma_5 u_{\mathrm{u}s1}(p_1)\,\overline{v}_{\mathrm{d}s2}(p_2)\gamma_5 v_{\mathrm{d}s2'}(p_2') \nonumber\\
&\quad+ 2\,\overline{u}_{\mathrm{u}s1'}(p_1')\gamma_5 v_{\mathrm{d}s2'}(p_2')\,\overline{v}_{\mathrm{d}s2}(p_2)\gamma_5 u_{\mathrm{u}s1}(p_1) \big\}\,,\label{eq:H_eff_NJL_pi_ori}
\end{align}
which is obtained from the NJL Lagrangian in the two flavor NJL model after the Legendre transform~\cite{Klimt:1989pm,Vogl:1989ea,Vogl:1991qt,Klevansky:1992qe}. The nonitalic and italic subscripts in the Dirac spinors, ${u_{\mathrm{f}s}(p)}$ and ${v_{\mathrm{f}s}(p)}$,  respectively represent the flavors and the spins. Meanwhile,  $p_1$ and $p_2$ are the momenta of the valence quark and the valence antiquark, respectively.
The coefficient $G_{\pi}$ is the coupling constant of the NJL interaction. In this interaction, only the combinations of Dirac bilinears relevant to the valence Fock sector LFWFs of the pion have been included with the instantaneous terms due to the NJL interactions neglected. The explicit expressions for the matrix elements of the NJL interactions within the BLFQ framework can be found in Ref.~\cite{Jia:2018ary}. 

To compute the Hamiltonian matrix we follow  BLFQ~\cite{Vary:2009gt} and  adopt the two-dimensional (2D) harmonic oscillator (HO) basis functions, to describe the transverse degrees of freedom, which are defined as~\cite{Li:2015zda}:
\begin{align}
	\phi_{nm}\left(\vec{q}_\perp;\kappa \right)& =\dfrac{1}{\kappa}\sqrt{\dfrac{4\pi n!}{(n+|m|)!}} \left(\dfrac{\vert\vec{q}_\perp\vert}{\kappa}\right)^{|m|} \exp\left(-\dfrac{\vec{q}_{\perp}^2}{2\kappa^2}\right)
	  L_n^{|m|} \left(\dfrac{\vec{q}_{\perp}^2}{\kappa^2}\right)\,e^{im\varphi}\,,\label{eq:def_phi_nm}
	\end{align}
	with $\tan(\varphi)=q_2/q_1$, $L_n^{|m|}(z)$ is the associated Laguerre polynomial, $n$ and $m$ are the radial and the angular quantum numbers, respectively. On the other hand, the basis functions in the longitudinal direction are defined as~\cite{Li:2015zda}
	\begin{align}
	&\quad \chi_l(x;\alpha,\beta)= \sqrt{4\pi(2l+\alpha+\beta+1)}\sqrt{\dfrac{\Gamma(l+1)\Gamma(l+\alpha+\beta+1)}{\Gamma(l+\alpha+1)\Gamma(l+\beta+1)}}  x^{\beta/2}(1-x)^{\alpha/2}\,P_l^{(\alpha,\beta)}(2x-1)\,,\label{eq:def_chi_l}
	\end{align}
	where $P_{l}^{(\alpha,\beta)}(z)$ is the Jacobi polynomial with the dimensionless parameters $\alpha$ $=$ $2m_{\overline{q}}(m_q+m_{\overline{q}})/\kappa^2$, $\beta$ $=$ $2m_q(m_q+m_{\overline{q}})/\kappa^2$, and $l=0,\,1,\,2,...$.
Using basis functions specified in Eqs.~(\ref{eq:def_phi_nm}) and (\ref{eq:def_chi_l}), the valence wave function in a given spin combination is expanded as 
\begin{align} 
	&\quad \psi_{rs}(x,\vec{k}_\perp) =\sum_{n, m, l}  \langle n, m, l, r, s | \psi\rangle~ \phi_{nm}\left(\dfrac{\vec{k}_\perp}{\sqrt{x(1-x)}};\kappa\right)\chi_l(x;\alpha,\beta),\label{eq:psi_rs_basis_expansions}
\end{align}
where the coefficients $\langle n, m, l,r,s|\psi\rangle$ represent the LFWFs in our BLFQ basis representation obtained from diagonalizing the truncated Hamiltonian.
We truncate the infinite dimensional Hilbert space of the valence Fock sector to a finite dimension by imposing the following restrictions on the quantum numbers~\cite{Jia:2018ary}:
	\begin{equation}
	0 \leq n \leq N_{\mathrm{max}}, \quad -2 \leq m \leq 2, \quad 0 \leq l \leq L_{\mathrm{max}}\,,
	\label{eq:nmax}
	\end{equation}
where $L_{\text{max}}$ specifies the basis resolution in the longitudinal direction, while $N_{\text{max}}$ controls the transverse momentum covered by 2D HO functions.
Since the NJL interactions do not couple to $\vert m\vert \geq 3$ basis states, we have a natural truncation for $m$~\cite{Jia:2018ary}.
%
The LFWFs of the mesons $\psi_{rs}(x,\vec{k}_\perp)$ are normalized as
\begin{align}
\sum_{r,s}
\int_0^1 \!\! \frac{dx}{2x(1-x)} \!\int \! \frac{d^2 \vec{k}_\perp}{(2\pi)^3}  \big | \psi_{rs}(x,\vec{k}_\perp)\big |^2 \!\! =\!\!1. \label{eq:normalization_LFWF}
\end{align}

Parameters in the BLFQ-NJL model have been fixed to generate the ground state masses of the light pseudoscalar as well as the vector mesons and the charge radii of the $\pi^+$ and the $K^+$~\cite{Jia:2018ary}. The LFWFs in this model provide a high quality description of the electromagnetic form factors \cite{Jia:2018ary}, PDFs for the pion and the kaon and pion-nucleus induced Drell-Yan cross sections~\cite{Lan:2019vui,Lan:2019rba}. The coefficient $\langle n, m, l, r, s \vert \psi \rangle$ for the neutral pion is assumed to be identical to that of the charged pion, resulting in the only difference between their wave functions being flavor.

%
%
%
%
%
\section{Parton distribution amplitude}\label{sec:pda}
%
The DAs are defined using the light-like separated gauge invariant vacuum-to-meson matrix elements. Explicitly in  the
light-front formalism, the leading-twist DAs $\phi_M(x;\mu)$ in the light-cone gauge for a pseudoscalar are defined by \cite{Lepage:1980fj,Brodsky:1997de}
\begin{align}
 \langle 0 | \overline\psi(z)\gamma^+\gamma_5\psi(-z)|M(p)\rangle_\mu =\,& \imag p^+ f_M\int_0^1 \dd x \, e^{\imag p^+z^-(x-\half)}
\phi_M(x;
\mu)\Big|_{z^+,\vec z_\perp=0}\,,
\end{align}
where $f_{M}$ are the decay constants. The non-local matrix elements as well as the DAs depend on the scale $\mu$, the renormalization scale which we take to be the UV
cutoff. Following these definitions, the DAs are normalized to unity:
\begin{equation}\label{norm}
 \int_0^1 \dd x \, \phi(x; \mu) = 1\,.
\end{equation} 
In terms of LFWF, the DAs of
pseudoscalar states can be written as \cite{Lepage:1980fj},
\begin{equation}\label{eq:da}
  \phi(x,\mu_0) =\frac{2\sqrt{2N_c}}{f_{M}} \frac{1}{\sqrt{x(1-x)}}
\int\limits \frac{\dd^2\vec{k}_\perp}{2(2\pi)^3}\psi_{\uparrow\downarrow\-\downarrow\uparrow}^{\lambda=0}(x, \vec
k_\perp)\,,
\end{equation}
where $\psi_{\uparrow\downarrow-\downarrow\uparrow} = (\psi_{\uparrow\downarrow}-\psi_{\downarrow\uparrow})/\sqrt{2}$ and $N_c$ is the number of colours.  In the BLFQ-NJL model, we use $N_c=3$. We compute the DA of the pion at the model scale using the LFWFs given in Eq.~(\ref{eq:psi_rs_basis_expansions}). Recall that the flavor wave function of the neutral pion is the only difference compared to the wave function of the charged pion in the BLFQ-NJL model. 

The QCD evolution of the DA is specified by the Efremov-Radyushkin-Brodsky-Lepage (ERBL) equations \cite{Efremov:1978rn,Efremov:1979qk,Lepage:1980fj}. In a Gegenbauer basis, one has \cite{RuizArriola:2002bp}
\begin{equation}\label{eq:evolDA}
	\phi(x,\mu)= 6x(1-x) \sum_{n=0}^{\infty} C_n^{3/2}(2x-1) a_n(\mu)\,,
\end{equation}
where $C_n^{\frac{3}{2}}(2x-1)$ is a Gegenbauer polynomial and
\begin{equation}
	a_n(\mu)=\frac{2}{3} \frac{2n+3}{(n+1)(n+2)} \left(\frac{\alpha_s(\mu)}{\alpha_s(\mu_{0})}\right)^{\gamma_n^{0}/2\beta_0} \int_0^1 \mathrm{d}x C_n^{3/2}(2x-1) \phi(x,\mu_0) \;,
\label{an}
\end{equation}
with
\begin{equation}
	\gamma_n^{(0)}=-2C_{\rm F} \left[3+ \frac{2}{(n+1)(n+2)}-4\sum_k^{n+1} \frac{1}{k} \right]; \quad \quad \beta_0=\frac{11}{3} C_{\rm A} -\frac{2}{3}n_{\rm f} \;,
\end{equation}
in the leading order (LO). Here, the color factors are given by $C_{\rm F}=\frac{4}{3}$ and $C_{\rm A}=3$. We take $n_{\rm f}=3$ to be the number of active flavors.
The strong running coupling is given by
\begin{equation}
	\alpha_{\rm s} (\mu) =\frac{4\pi}{\beta_0\ln \left(\mu^2/\Lambda^2_{\mathrm{QCD}}\right)} \,,
\end{equation}
with $\Lambda_{\mathrm{QCD}}=0.204$ GeV being the QCD scale parameter.
It is also useful to compute the moments in order to quantitatively compare with other theoretical predictions. The $p$-th
moment of the DA is defined as,
\begin{align}
\langle z_p \rangle = \int_0^1 dx~z^p~\phi(x,\mu)\,,
\end{align}
where $z\equiv(2x-1)$ when $p\ge 1$ and $z\equiv x$ for $p = -1$. Meanwhile, the moments of the DA are directly related to the Gegenbauer coefficients in Eq.~(\ref{eq:evolDA}). The second, fourth and sixth moments are expressed as \cite{Choi:2007yu}
\begin{equation}
	\langle z_2 \rangle = \frac{12}{35} a_2 + \frac{1}{5} \;,
\end{equation}
\begin{equation}
	\langle z_4 \rangle = \frac{3}{35} + \frac{8}{35} a_2 + \frac{8}{77} a_4 \;,
\end{equation}
\begin{equation}
	\langle z_6 \rangle = \frac{1}{21} + \frac{12}{77} a_2 + \frac{120}{1001} a_4+ \frac{64}{2145} a_6 \;.
\end{equation}

\begin{figure*}
	\centering
	\includegraphics[width=7.5cm]{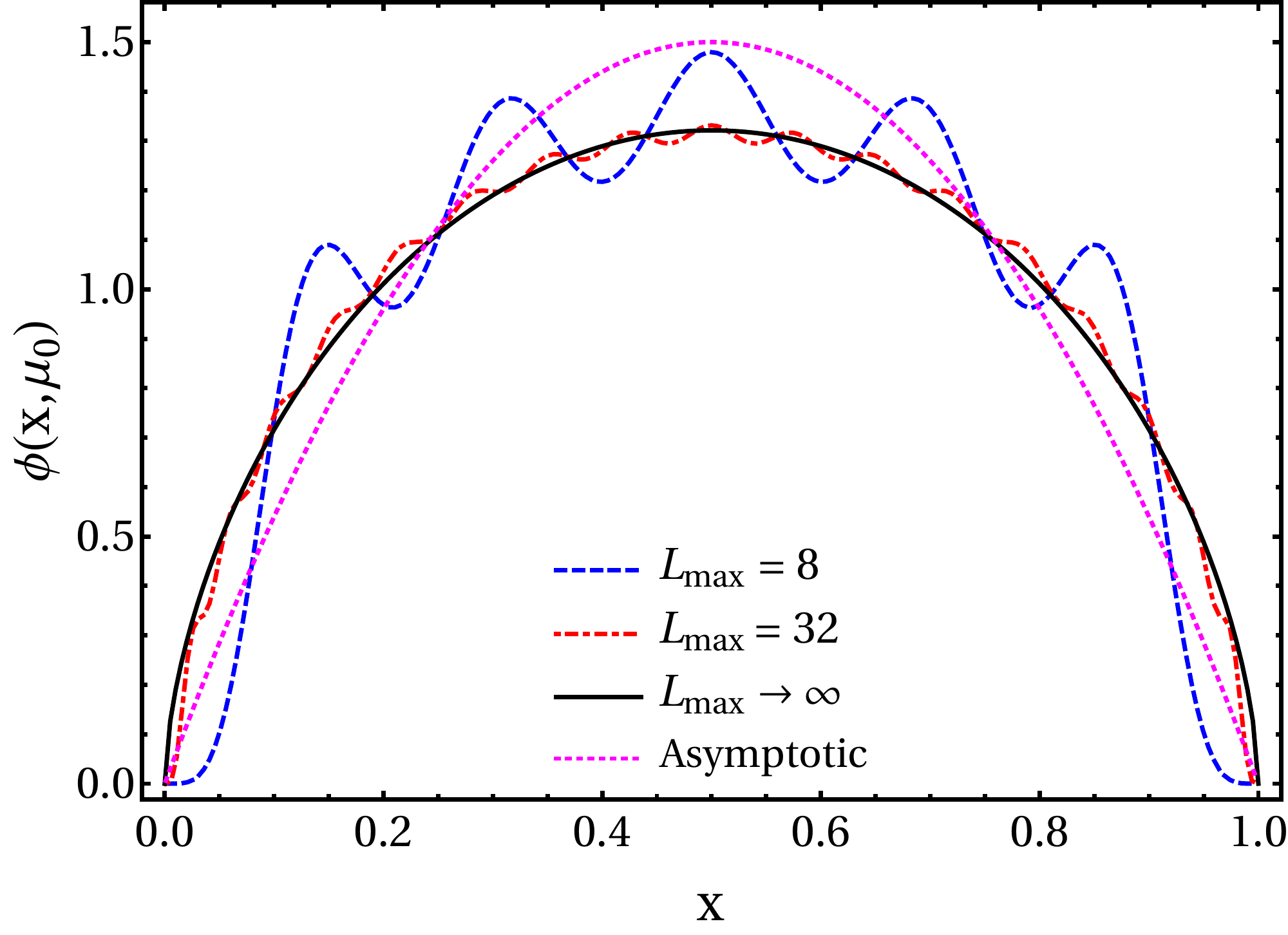}
	\hspace{0.1cm}
	\includegraphics[width=7.5cm]{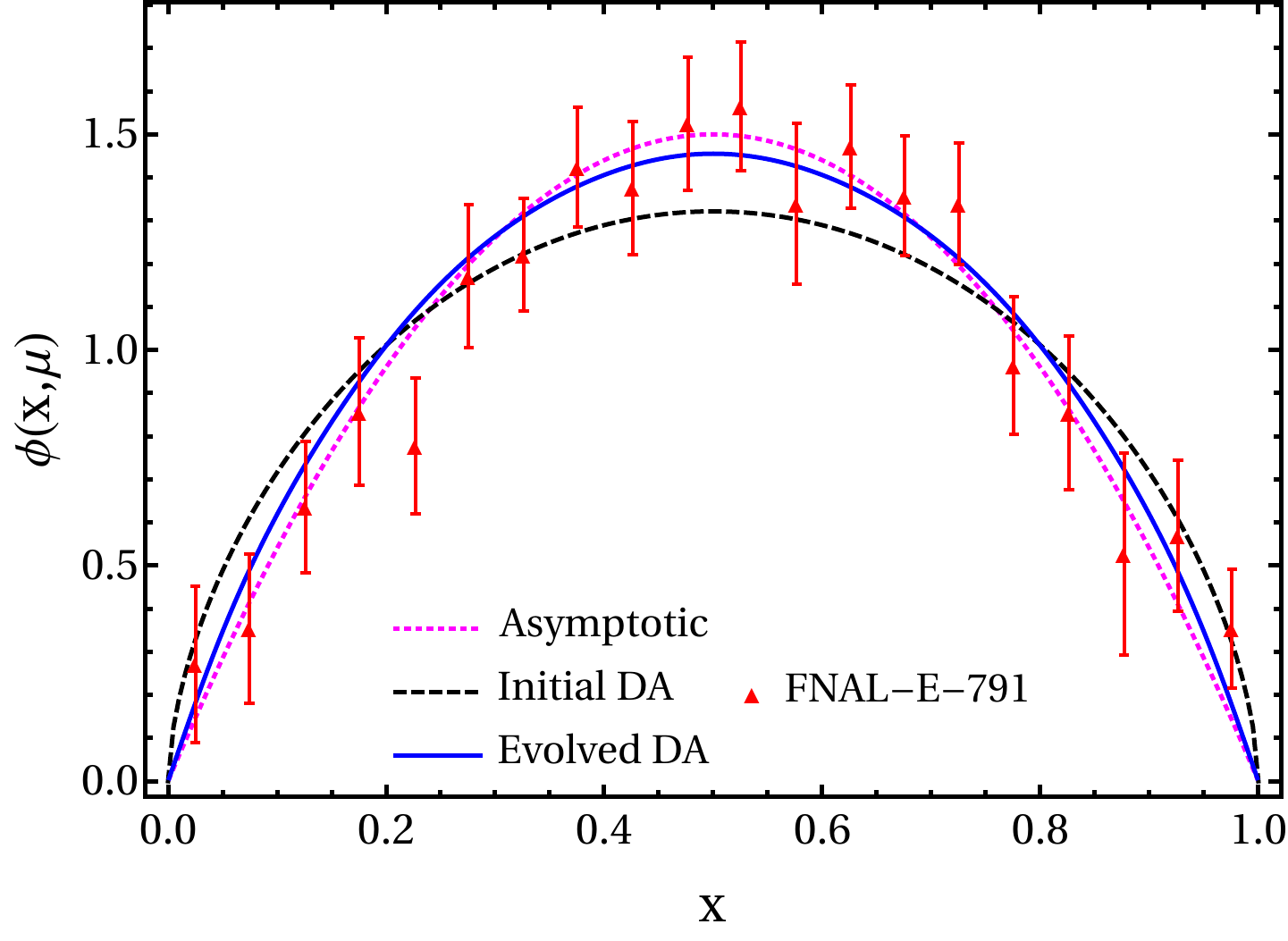}
	\caption{The valence quark DA of the pion in the BLFQ-NJL model. Left panel: the DAs for $L_{\rm max}=8$ (blue dashed) $L_{\rm max}=32$ (red dash-dotted) and extrapolated to $L_{\rm max}\to\infty$ that fits to Eq.~\eqref{eq:PDA_fits} with $a=b=0.62$ (black-solid) at the model scale. Right panel: the evolved DA from the initial scale ($\mu_0^2=0.120\pm0.012$ GeV$^2$) using ERBL equations to the experimental scale of $10$ GeV$^2$. The black dashed line corresponds to the DA at the initial scale, while the blue solid line represents the evolved DA. Our result is compared with the FNAL-E-791 data~\cite{Aitala:2000hb} and the asymptotic DA (magenta dotted): $6x(1-x)$~\cite{RuizArriola:2002bp}.}
	\label{fig:DA_initial}
\end{figure*}

The LFWFs of the valence quarks in the light mesons have been solved in the BLFQ framework using the NJL interactions as discussed in the Section~\ref{sc:BLFQ_NJL}. We employ the wave functions obtained with truncated basis to compute the DA of the pion using Eq.~(\ref{eq:da}). On the left panel of Fig.~\ref{fig:DA_initial}, we show the valence quark DA of the pion at the model scale using the basis truncation $N_\text{max} = 8$ with $L_\text{max} =8,\, 32$, and extrapolation to $\infty$. 
The oscillations  on the DA are numerical artifacts, while with increasing $L_\text{max}$ the DA tends toward a smooth function with decreasing oscillation amplitude about a single-peaked function. The DA for $L_\text{max}\to\infty$ is therefore fitted to the following functional form~\cite{Jia:2018ary}:
\begin{equation}
\phi(x,\mu_0)=\dfrac{x^a (1-x)^b}{B(a+1,b+1)} \label{eq:PDA_fits}\,,
\end{equation}
with $a=b=0.6$, where $B(a+1,b+1)$ is the Euler beta function. We illustrate the pion valence quark DA after QCD evolution on the right panel of Fig.~\ref{fig:DA_initial}. Explicitly,
we evolve our input DA from the model scale $\mu_0^2=0.120\pm0.012$ GeV$^2$~\cite{Lan:2019rba} to the experimental scale of the  FNAL-E-791 experiment, $10$ GeV$^2$~\cite{Aitala:2000hb}. The initial scale was determined by requiring the PDF result after QCD evolution utilizing the LO DGLAP equation to fit the pion valence quark PDF data from the FNAL-E-615 experiment~\cite{Conway:1989fs}. As can be seen, our evolved DA is in excellent agreement with the FNAL-E-791 data. On the other hand, the pion DA in the BLFQ-NJL model is very close to the asymptotic DA already at $\mu^2=10$ GeV$^2$.

\begin{table*}
\caption{Our predictions for the first three non-vanishing moments and inverse moment of the pion DA, compared to other theoretical predictions.}\label{tab:moments}
\centering
\begin{tabular}{lccccccc}
\hline\hline
& ~$\mu{\rm [GeV]}$~ & ~$\langle z_2\rangle$~ & ~$\langle z_4\rangle$~ & ~$\langle z_6\rangle$~ & ~$\langle x^{-1}\rangle$ \\ \hline
BLFQ-NJL (this work)             & 1, 2  & 0.221, 0.217 & 0.099, 0.097 & 0.057, 0.055 & 3.21, 3.17\\
Playkurtic~\cite{Stefanis:2014nla}         & 2  & $0.220^{+0.009}_{-0.006}$ & $0.098^{+0.008}_{-0.005}$   &-& $3.13^{+0.14}_{-0.10}$\\
NLC Sum Rules~\cite{Bakulev:2001pa}      & 2  & $0.248^{+0.016}_{-0.015}$ & $0.108^{+0.05}_{-0.03}$     &-& 3.16(9)\\
LF Quark Model~\cite{Choi:2007yu}     & $\sim 1$ & 0.24(22) & 0.11(9)                                & 0.07(5)&-\\
Sum Rules~\cite{Ball:2004ye}          & 1  & 0.24 & 0.11                                             &-&-\\
AdS/QCD~\cite{Brodsky:2011yv}  &$\sim 1$ & 0.25 & 0.125                                      & 0.078 & 3.98\\
LF Holographic ($B=0$)~\cite{Ahmady:2018muv}  &1, 2 & 0.180, 0.185 & 0.067, 0.071                             &-& 2.81, 2.85\\
LF Holographic ($B\gg1$)~\cite{Ahmady:2018muv} &1, 2 & 0.200, 0.200 & 0.085, 0.085                             &-& 2.93, 2.95\\
Renormalon model~\cite{Agaev:2005rc}   & 1  & 0.28 & 0.13                                             &-&-\\
Instanton vacuum (MIA 1)~\cite{Nam:2006au}& 1, 2  & 0.237, 0.218  & 0.112, 0.094                                 & 0.066, 0.052 &-\\
Instanton vacuum (MIA 2)~\cite{Nam:2006au}& 1, 2 & 0.239, 0.220  & 0.113, 0.096                                 & 0.067, 0.053 &-\\
Sum Rules~\cite{Chernyak:1983ej}          & 2  & 0.343  & 0.181                                          &-& 4.25  \\
Dyson-Schwinger [RL, DB]~\cite{Chang:2013pq}  & 2 & 0.280, 0.251& 0.151, 0.128                          &-& 5.5, 4.6 \\
QCD background field theory sum rule~\cite{Zhong:2021epq}               & 1  & 0.271(13) & 0.138(10)& 0.087(6)& 3.95\\
QCD background field theory sum rule~\cite{Zhong:2021epq}               & 2  & 0.254(10) & 0.125(7) & 0.077(6)& 3.33\\
Lattice QCD~\cite{Arthur:2010xf}            & 2  & 0.28(1)(2) & -                                          &-&-\\
Lattice QCD~\cite{Braun:2015axa}           & 2  & 0.2361(41)(39)                & -                       &-&-\\
Lattice QCD~\cite{Braun:2006dg}          & 2  & 0.27(4)                       & -                       &-&-\\
Lattice QCD~\cite{Bali:2017ude}          & 2  & 0.2077(43)                       & -                       &-&-\\
Lattice QCD~\cite{Bali:2019dqc}          & 2  & 0.234(6)(6)                       & -                       &-&-\\
Lattice QCD~\cite{Zhang:2020gaj}          & 2  & 0.244(30)                       & -                       &-&-\\ 
Asymptotic QCD             &$\infty$& 0.200 & 0.086    & 0.048    & 3.00\\
\hline\hline
\end{tabular}
\end{table*}

The numerical values of the first three non-vanishing moments and the inverse moment of the pion DA in the BLFQ-NJL model are presented in Table~\ref{tab:moments}. We compare our predictions with the results obtained from various theoretical approaches. Our predictions for $z_2$ and/or $z_4$ roughly agree with those in Refs.~\cite{Stefanis:2014nla,Bakulev:2001pa,Choi:2007yu,Ball:2004ye,Brodsky:2007hb,Ahmady:2018muv,Nam:2006au,Braun:2015axa,Bali:2017ude,Bali:2019dqc,Zhang:2020gaj}. For $z_6$, the numerical values are compared with Refs.~\cite{Choi:2007yu,Brodsky:2011yv,Nam:2006au,Zhong:2021epq}, with our predictions being close to the results evaluated from the QCD instanton vacuum~\cite{Nam:2006au}. The inverse moment of the pion DA in the BLFQ-NJL model is in good agreement with Refs.~\cite{Stefanis:2014nla,Bakulev:2001pa}, while differing from other predictions summarized in Table~\ref{tab:moments}.

Based on the BLFQ-NJL model, the decay constant for the pion is given in Table~ \ref{tab:decay_constant}. In order to gain a clear impression on how basis truncation
affects the decay constant, we include the numerical values with basis cutoffs $N_{\rm max}=8$ with $L_{\rm max}=8$, $16$, and $32$ to demonstrate a good
convergence trend. With $L_{\rm max}=32$ we predict $f_\pi=145.3$ MeV, while the experimental data is $f_\pi^{\rm exp}= 130.2$ MeV~\cite{ParticleDataGroup:2018ovx}. Note that the decay constants computed here differ from the results presented in Ref.~\cite{Jia:2018ary}, where an overall factor ($\sqrt{2}$) was erroneously included.


We now use the Alder-Bell-Jackiw (ABJ) anomaly relations \cite{Adler:1969gk,Bell:1969ts} to compute the pion-photon TFF at zero momentum transfer as follows \cite{Choi:2017zxn,Ahmady:2018muv}:
\begin{equation}
	F_{\pi \gamma}^{\mathrm{ABJ}}(0)=\frac{1}{2\sqrt{2}\pi^2f_{\pi^0}} \;,
\end{equation}
so that we can evaluate the radiative decay width using
\begin{equation}
	\Gamma_{\pi \to \gamma \gamma}= \frac{\pi}{4}\alpha^2_{\mathrm{EM}} M_\pi^3 |F_{\pi\gamma}(0)|^2 .
\end{equation}
Our results are presented in Table~\ref{tab:decay_constant}, where we find reasonable  agreement with the experimental value~\cite{ParticleDataGroup:2018ovx}. With the largest basis size in our current calculation, i.e., $N_{\rm max}=8$ and $L_{\rm max}=32$, we obtain $\Gamma_{\pi \to \gamma \gamma}=6.98\times10^{-3}$ keV, whereas the experimentally measured value is $\Gamma_{\pi \to \gamma \gamma}=7.82\times10^{-3}$ keV. 

With the original model parameters given in Ref.~\cite{Jia:2018ary}, our prediction  for the pion decay constant is somewhat larger than the experimental data, which effectively leads to a slightly smaller radiative decay width compared to the experimental value. However, if we relax the constraint from the charge radius, we can modify the model parameters into $m_q=356.77$ MeV, $\kappa=182.35$ MeV, and $G_\pi=2.6573\times 10^{-5}$ MeV for the basis cutoff $N_{\rm max}=8$ and $L_{\rm max}=32$ in order to reproduce the experimental decay constant.

\begin{table*}
\caption{Our predictions for the decay constant, $f_\pi$, of the pion and the radiative decay width, $\Gamma_{\pi\to 2\gamma}$, compared to the measured values from
Particle Data Group (PDG)~\cite{ParticleDataGroup:2018ovx}. The BLFQ-NJL model results are quoted using the LFWFs at the basis cutoff $N_{\rm max}=8$ and $L_{\rm max}=8$, $16$ and $32$, respectively.}
\label{tab:decay_constant}
\centering
\begin{tabular}{cccc|c} \hline\hline
$[N_{\rm max},\,L_{\rm max}]\to$ ~~&~~ $[8, 8]$ ~~&~~ $[8, 16]$ ~~&~~ $[8, 32]$ ~~&~~ Experimental data~\cite{ParticleDataGroup:2018ovx} \\
\hline
 $f_\pi$ (MeV) & 142.8 & 144.8 & 145.3 & 130.2\\
 $\Gamma_{\pi\to 2\gamma}$ (keV) & 7.22 $\times$ 10$^{-3}$ & 7.03 $\times$ 10$^{-3}$ & 6.98 $\times$ 10$^{-3}$ & (7.82 $\pm$ 0.22) $\times$ 10$^{-3}$ \\
\hline\hline
\end{tabular}
\end{table*}

\section{Pion to photons transition form factors}\label{sec:TFF}
The meson-photon transition form factor, $F_{M\gamma}(Q^2)$, of a pseudoscalar meson ($M$) for the $M\to\gamma^*\gamma$ decay is defined through
the matrix element of electromagnetic current as:~\cite{Lepage:1980fj}
\begin{align}\label{Eq:TFF}
\langle\gamma(P-q)|J^\mu|M(P)\rangle= -i e^2 F_{M\gamma}(Q^2)\epsilon^{\mu\nu\rho\sigma}P_\nu\epsilon_\rho q_\sigma\,,
\end{align}
where $P$ and $q$ are the momenta of the meson and the virtual photon, respectively.  The invariant 4-momentum transfer squared $Q^2=-q^2>0$ and $\epsilon_\rho$ is the transverse polarization vector of the final (on-shell) photon. The TFF, $F_{M\gamma}(Q^2)$, can be evaluated from the convolution of a hard scattering amplitude (HSA), $T_{\mathrm H}(x,Q^2)$, calculable in perturbation theory, with a nonperturbative DA~\cite{Lepage:1980fj,Brodsky:1981rp,Brodsky:2011yv} using
\begin{align}
Q^2F_{\pi \gamma}(Q^2)=\frac{\sqrt{2}}{3}f_\pi \int_0^1 {\rm d}x \,
T_{\mathrm H}(x,Q^2) \, \phi(x,(1-x) Q),
\label{eq:TFF_convo}
\end{align} 
where $T_{\mathrm H}(x,Q^2)$ to the next-to-leading order (NLO) is given by~\cite{delAguila:1981nk,Braaten:1982yp,Kadantseva:1985kb,Melic:2001wb,Melic:2002ij},
\begin{align}
T_{\mathrm H}(x,Q^2)&=\frac{1}{1-x}
+\frac{\alpha_s(\mu_{\mathrm R})}{4 \pi}
C_{\mathrm F} \frac{1}{1-x} \,\Bigg[-9-\frac{1-x}{x} {\rm ln}(1-x) \nonumber\\
& + {\rm ln}^2(1-x)   +\left\{ 3+2 \,{\rm ln}(1-x) \right\} {\rm ln} \left(\frac{Q^2}{\mu^2_{\mathrm R}} \right)
\Bigg].
\label{eq:THNLO}
\end{align}
For simplicity, the regularization scale, $\mu_{\mathrm R}^2$, is taken as $\mu_{\mathrm R}^2=Q^2$ to eliminate large logarithm terms. At leading order, only the first term in Eq.~(\ref{eq:THNLO}) contributes to the TFF. Using the standard hard scattering approach, the  NLO corrections have been studied in Refs.~\cite{delAguila:1981nk,Braaten:1982yp,Kadantseva:1985kb,Melic:2001wb,Melic:2002ij} under the assumption that ${\phi(x,(1-x)Q) \simeq \phi(x, Q)}$. The replacement is reasonable for the behavior at the asymptotic limit $Q^2\to \infty$. However, this approximation is not well justified for the calculation below the asymptotic region, where one needs to take into account the ERBL evolution effects. A proper treatment of the NLO calculations with the evolution effects and $\phi(x,(1-x) Q)$ has been illustrated in Ref.~\cite{Brodsky:2011yv}.

\begin{figure*}
	\centering
	\includegraphics[width=7.5cm]{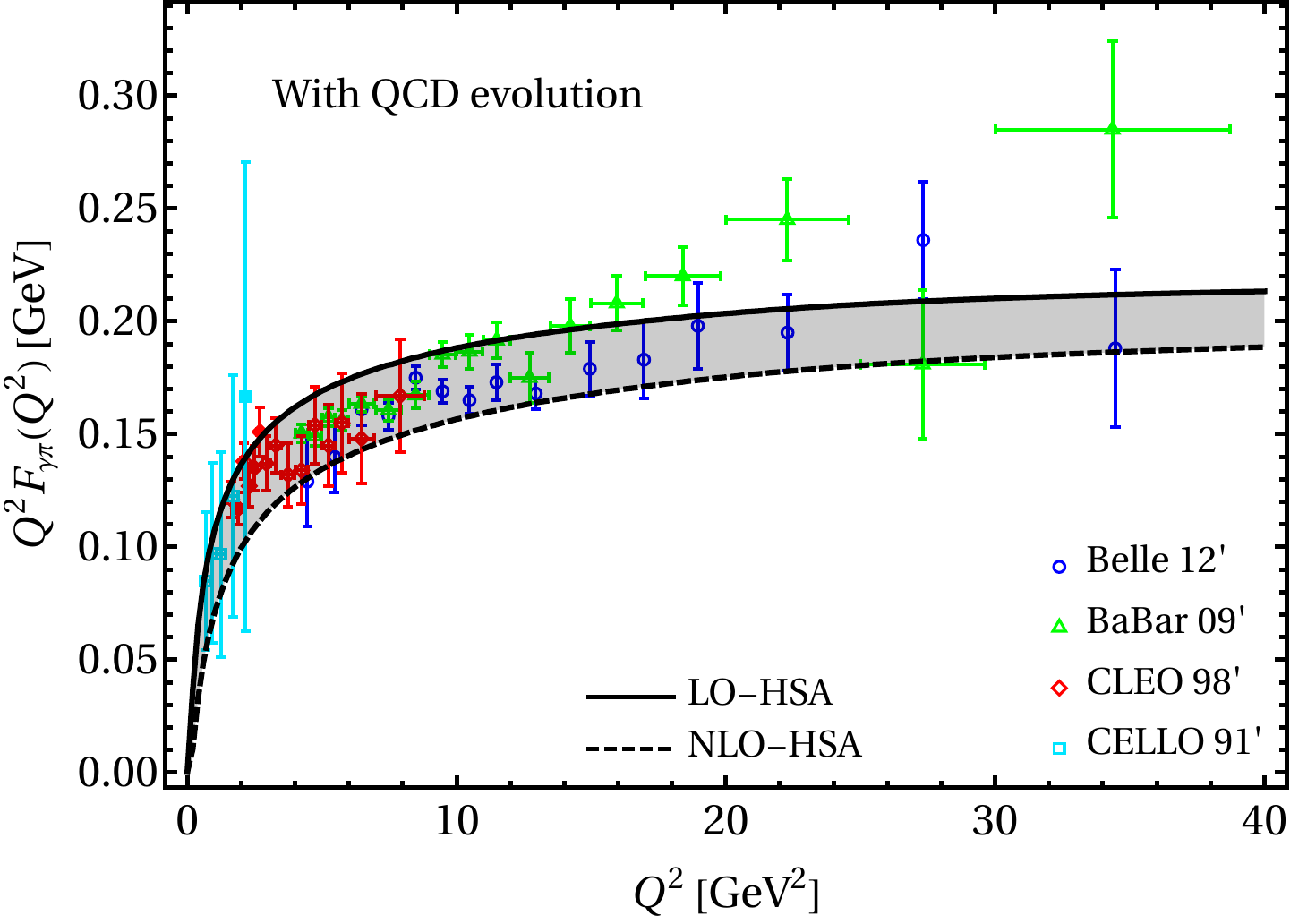}
	\includegraphics[width=7.5cm]{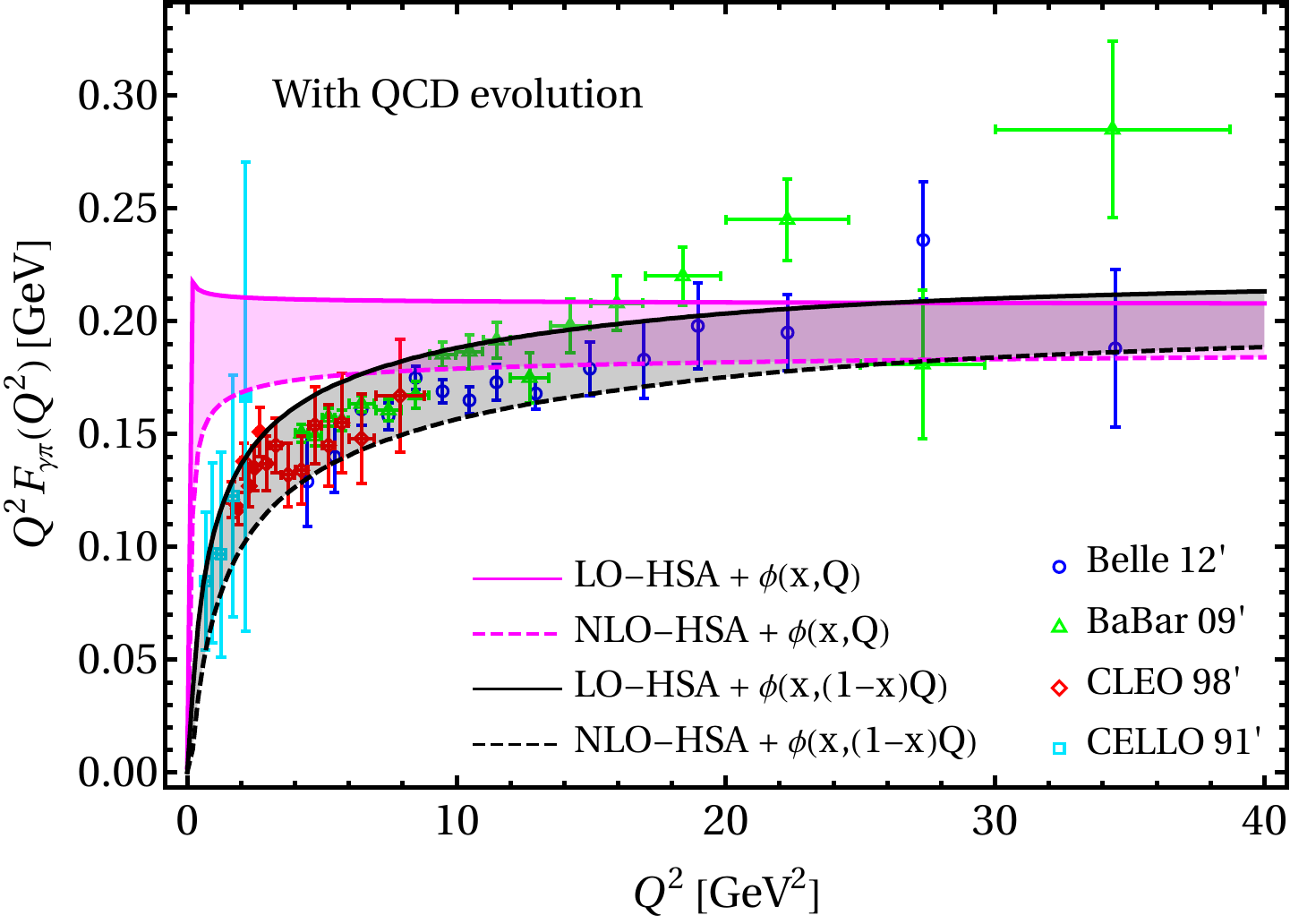}
	\caption{The $\pi^0\to \gamma^*\gamma$ transition form factor with BLFQ basis truncation $N_{\rm max}=8$ and $L_{\rm max}=32$. Left panel: the solid and dashed lines correspond to the results with LO and NLO hard scattering amplitudes, respectively. Right panel: Comparison of the TFFs evaluated with the DAs $\phi(x, Q)$ (magenta lines) and $\phi(x,(1-x)Q)$ (black lines) in Eq.~(\ref{eq:TFF_convo}). Effect of the scale evolution is considered following Eq.~(\ref{eq:evolDA}).}
	\label{fig:TFF1}
\end{figure*}
\begin{figure*}
	\centering
	\includegraphics[width=7.5cm]{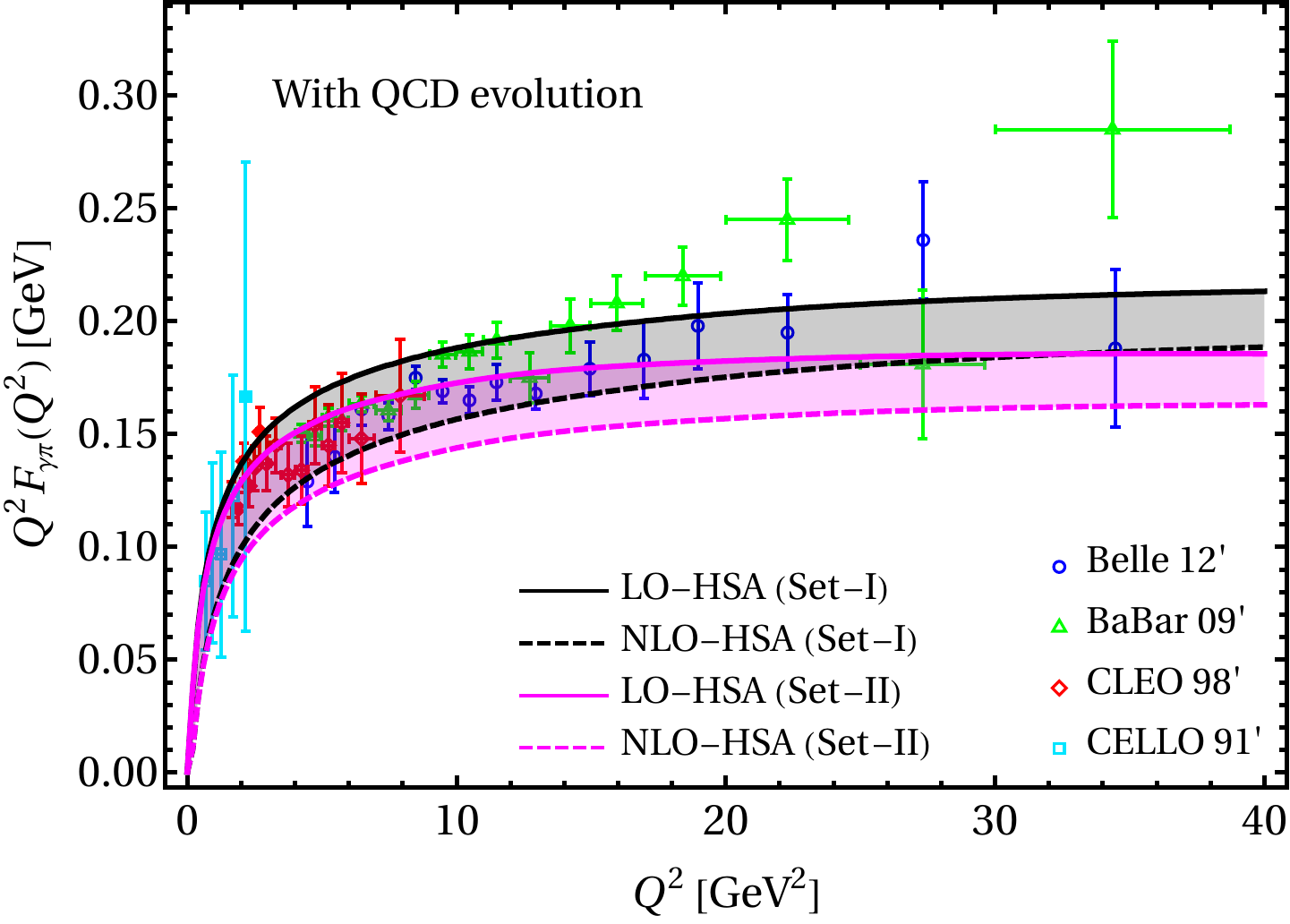}
	\caption{Comparison of the $\pi^0\to \gamma^*\gamma$ transition form factors evaluated with the original model parameters (Set-I) given in Ref.~\cite{Jia:2018ary} (black lines) and the re-adjusted parameters (Set-II) that fit the pion experimental pion decay constant (magenta lines) both with BLFQ basis truncation $N_{\rm max}=8$ and $L_{\rm max}=32$.}
	\label{fig:parameter}
\end{figure*}
\begin{figure*}
	\centering
	\includegraphics[width=7.5cm]{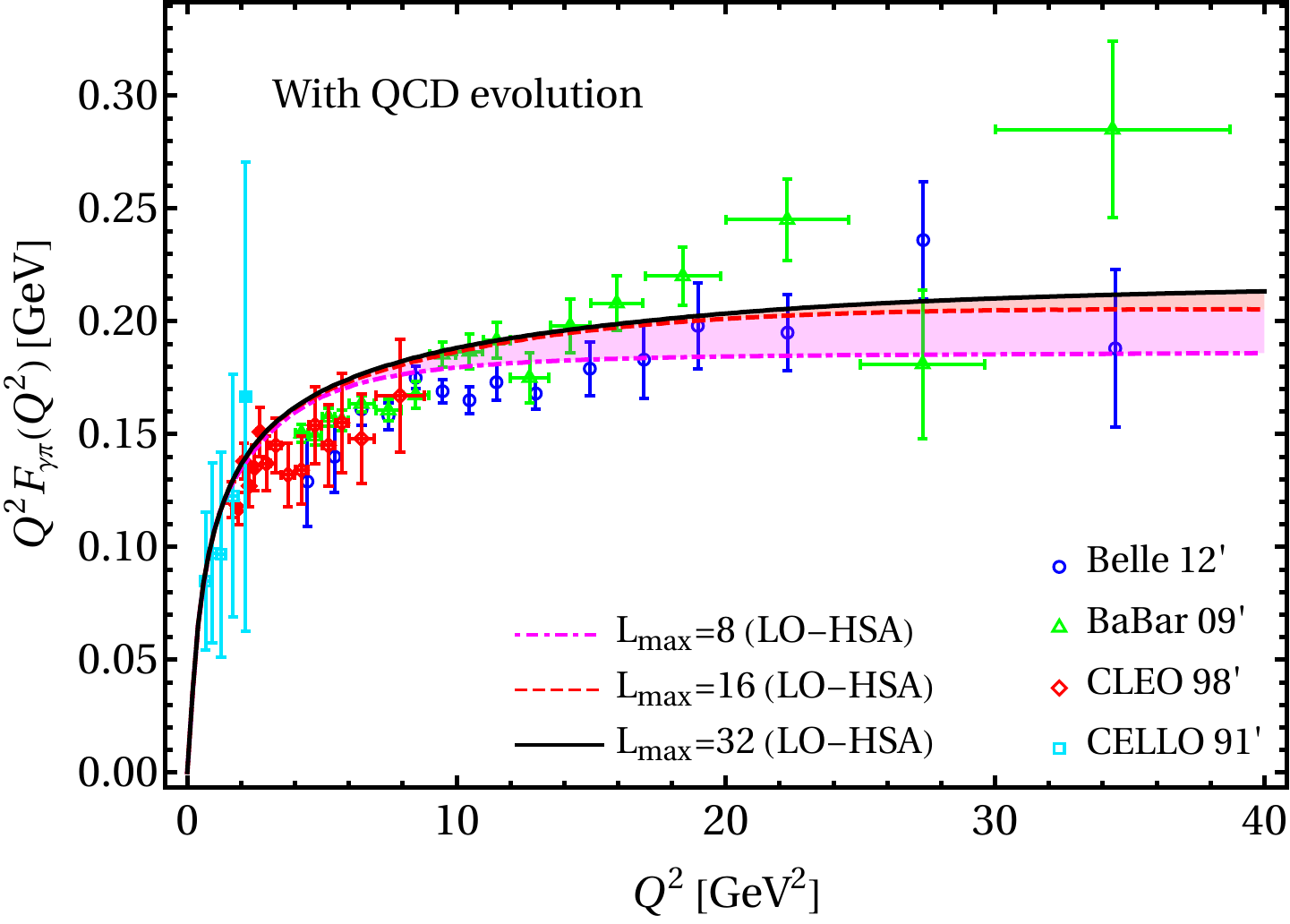}
	\includegraphics[width=7.5cm]{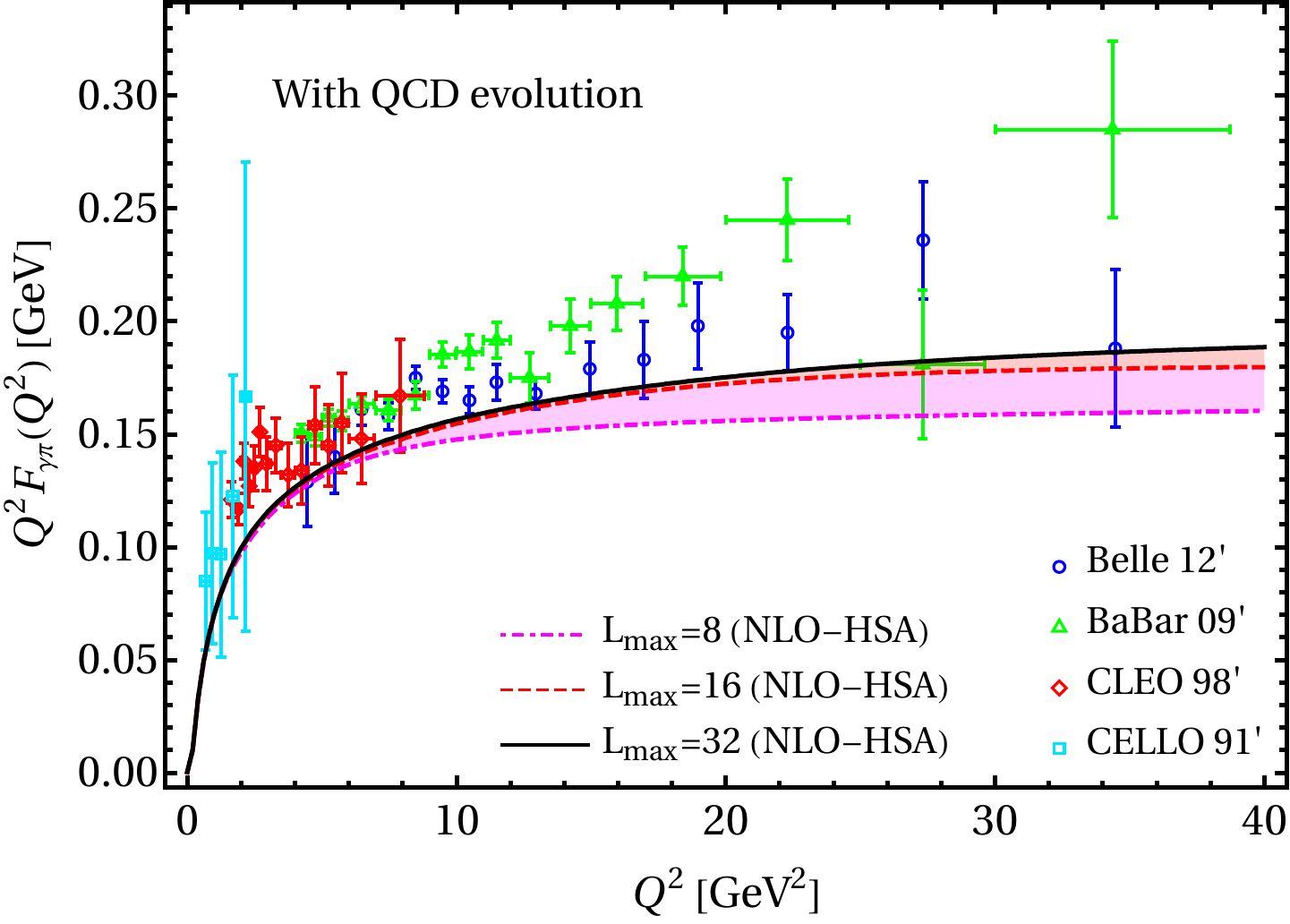}
	\caption{Sensitivity of the $\pi^0\to \gamma^*\gamma$ transition form factor to BLFQ basis truncation. The magenta dot-dashed, red dashed, black solid lines correspond to the $\{N_{\rm max},\, L_{\rm max}\}\equiv \{8,\,8\}$, $\{8,\,16\}$, and $\{8,\,32\}$, respectively. Left and right panels represent the results with LO and NLO hard scattering amplitude, respectively.}
	\label{fig:TFF3}
\end{figure*}

Inserting the evolved DA from Eq.~(\ref{eq:evolDA}) into Eq.~(\ref{eq:TFF_convo}) we evaluate $Q^2F_{\pi \gamma}(Q^2)$ with the ERBL evolution considered.  Figure~\ref{fig:TFF1} (left panel) shows our results for the pion to photon TFF, $Q^2F_{\pi \gamma}(Q^2)$, in the BLFQ-NJL model and compares with the available experimental data from the Belle~\cite{Uehara:2012ag}, the BaBar~\cite{Aubert:2009mc}, the CLEO~\cite{CLEO:1997fho}, and the CELLO~\cite{CELLO:1990klc} Collaborations. The gray band corresponds to the correction in the TFF due to the $\alpha_s$ order correction in the HSA in Eq.~(\ref{eq:THNLO}). We need to set $(1-x)Q=\mu_0$ for $ (1-x) Q<\mu_0$ to ensure the convergence of the integration in Eq.~(\ref{eq:TFF_convo}). We obtain the TFF results with the BLFQ basis truncation: $N_{\rm max}=8$ with $L_{\rm max}=32$ and find good agreement of our calculated pion TFF with the experiments performed by the Belle~\cite{Uehara:2012ag}, the CLEO~\cite{CLEO:1997fho}, and the CELLO~\cite{CELLO:1990klc} Collaborations. However, it deviates from the rapid growth in the large $Q^2$ region reported by the BaBar Collaboration~\cite{Aubert:2009mc}. There are also theoretical studies suggesting that the BaBar data are incompatible with QCD calculations~\cite{Mikhailov:2009sa,Roberts:2010rn,Bakulev:2011rp,Wu:2011gf}. It has also been demonstrated in Ref.~\cite{Brodsky:2011yv} that the explanation of BaBar data at large $Q^2$ with the QCD calculations
using the asymptotic QCD, AdS/QCD, and Chernyak-Zhitnitsky models for the pion DA is not possible, but can be accommodated by a flat modeling of the pion DA~\cite{Polyakov:2009je,Brodsky:2011yv}. However, the calculations with such a DA underestimate
significantly the pion TFF at low $Q^2$. Such a DA also shows a trend that would appear to violate the Brodsky-Lepage limit of $Q^2 F_{\pi \gamma}(Q^2 \to \infty)=2 f_\pi$~\cite{Brodsky:2011yv}. Except those using the flat modeling of the pion DA~\cite{Dorokhov:2009dg,Radyushkin:2009zg,Polyakov:2009je,Li:2009pr}, there exist phenomenological studies reproducing the BaBar data for the pion to photon TFF~\cite{Wu:2010zc,Kroll:2010bf,RuizArriola:2010mrb,Gorchtein:2011vf,Pham:2011zi,Dorokhov:2010zzb,Agaev:2010aq,Kotko:2009ij}.

In Fig.~\ref{fig:TFF1} (right panel), we compare the results calculated using $\phi(x,{ (1-x)} Q)$ and $\phi(x,Q)$ in Eq.~(\ref{eq:TFF_convo}). We notice that it is a reasonable approximation to use $\phi(x,Q)$ in the perturbative regime, where both $\phi(x,(1-x) Q)$ and $\phi(x,Q)$ lead to almost identical pion TFF. However, using $\phi(x,Q)$ fails to reproduce the experimental data at low $Q^2$ region, particularly for the $Q^2<10$ GeV$^2$, whereas the results using $\phi(x,(1-x) Q)$ agree well with the available data.

Parameters in the BLFQ-NJL model have been fixed to fit the charge radius of the pion~\cite{Jia:2018ary}, while they lead to a slightly larger pion decay constant compared to the experimental data as summarized in Table~\ref{tab:decay_constant}. However, the experimental the decay constant can be obtained by adjusting the model parameters into those  mentioned by the end of Section~\ref{sec:pda}. In order to gain a clear impression on how the change in parameters affects $Q^2F_{\pi \gamma}(Q^2)$, in Fig.~\ref{fig:parameter} we present a comparison of the singly virtual pion to two photon TFF computed using the original model's parameters given in Ref.~\cite{Jia:2018ary} and the modified parameters. We find that the two sets of parameters provide qualitatively similar results for this TFF at small $Q^2$. However, $F_{\pi \gamma}(Q^2)$ with new parameters is narrower than that obtained using the original model's parameters, which accounts for the difference in the TFF for large $Q^2$.


The sensitivity of the BLFQ-NJL model prediction to basis truncation is shown in Fig.~\ref{fig:TFF3}, where we present the results for $\{N_{\rm max},\, L_{\rm max}\}\equiv \{8,\,8\}$, $\{8,\,16\}$, and $\{8,\,32\}$. The results show a good
convergence trend over the range of $Q^2$ as evident by finding
that the $L_{\rm max}=16$ and $L_{\rm max}=32$ results
nearly coincide with each other in contrast to the $L_{\rm max}=8$ results presented in Fig.~\ref{fig:TFF3}. This observed
convergence in the TFFs is reassuring since the DAs are also reasonably well converged as can be seen in the left panel of Fig.~\ref{fig:DA_initial}. The
difference between the $L_{\rm max}=32$ and $8$ values is
presented as our uncertainty estimate in the pion TFF.

We now turn our attention to the case in which the photons as decay products 
are both off mass-shell, i.e., for the TFF $F_{\pi\gamma^*}(Q_1^2,Q_2^2)$. This pion to doubly virtual photons ($\pi^0\to \gamma^* \gamma^*$) TFF can be obtained by replacing the hard-scattering amplitude $T_{\mathrm H}$ in the previous analysis with an appropriate  expression. The TFF, $F_{M\gamma^*}(Q_1^2,Q_2^2)$, can be expressed from the convolution of a HSA, $T_{\mathrm H}(x,Q_1^2,Q_2^2)$ with a nonperturbative DA~\cite{Lepage:1980fj,Brodsky:1981rp,Brodsky:2011yv},
\begin{align}
F_{\pi \gamma^*}(Q_1^2,Q_2^2)=\frac{\sqrt{2}}{3}f_\pi \int_0^1 {\rm d}x \,
T_{\mathrm H}^{\gamma^* \gamma^* \to \pi^0}(x,Q_1^2,Q_2^2) \, \phi(x,\bar{Q}),
\label{eq:DTFF_convo}
\end{align} 
where we assume $\bar{Q}=(1-x)Q_1+xQ_2$ and at the leading order $T_{\mathrm H}$ has the form
\begin{align}
T_{\mathrm H}^{\gamma^* \gamma^* \to \pi^0}(x,Q_1^2,Q_2^2)=\frac{1}{ (1-x) Q_1^2 + x Q_2^2}.
\label{eq:DTH}
\end{align} 
Note that the singly virtual TFF $F_{\pi \gamma}(Q^2)$ can be obtained by setting one of the momentum transfers to zero in Eq.~(\ref{eq:DTFF_convo}). While the $F_{\pi \gamma}(Q^2)$ 
is sensitive to the shape of the pion DA, the doubly virtual TFF $F_{\pi \gamma^*}(Q_1^2,Q_2^2)$ when $Q_1^2 \neq 0$ and $Q_2^2 \neq 0$ is much less sensitive to the end-point behavior of the pion DA, since the HSA in Eq.~(\ref{eq:DTH}) is well-behaved at the end-points, $x=\{0,1\}$. It can also be noted that, for the kinematic region satisfying $Q_1^2=Q_2^2$, the HSA becomes independent of $x$ and thus $Q_1^2F_{\pi \gamma^*}(Q_1^2,Q_1^2)$ only depends on the normalization of the pion DA, which is scale and model independent.
One then has $Q^2F_{\pi \gamma^*}(Q^2,Q^2)\to \frac{\sqrt{2}}{3}f_\pi$.
 For this transition to two off-shell photons, the pQCD
expression of $T_{\mathrm H}^{\gamma^* \gamma^* \to \pi^0}$ at NLO can be found in Ref.~\cite{Braaten:1982yp}.

\begin{table*}
\caption{The  transition form factors $F_{\pi\gamma^*}(Q^2_1,Q^2_2)$ (in units of $10^{-3}$ GeV$^{-1}$) for some ($Q^2_1,Q^2_2$) values (in units of GeV$^2$) compared with the LFQM~\cite{Choi:2019wqx}, the vector meson dominance (VMD) model~\cite{Choi:2019wqx}, and the pQCD predictions. The first and second values  for each row entry  in the second and third columns correspond to the basis truncations $L_{\rm max}=8$ and $32$, respectively. The
difference between the $L_{\rm max}=8$ and $32$ values are
presented as our uncertainty estimate.}
\label{tab:DTFF}
\centering
\begin{tabular}{ccccccc} \hline\hline
 $(Q^2_1,Q^2_2)$&$F^{\rm LO}_{\pi\gamma^*}$&$F^{\rm NLO}_{\pi\gamma^*}$&pQCD&pQCD &LFQM~\cite{Choi:2019wqx}&VMD~\cite{Choi:2019wqx}\\
 &(this work)&(this work)&LO&NLO&&  \\
\hline
 (6.48, 6.48)& 10.39 - 10.56 & 9.59 - 9.75 & 9.52 & 8.78 & 9.08 & 1.957 $\pm$ 0.022\\
 (16.85, 16.85)& 3.99 - 4.06 & 3.73 - 3.79 & 3.66 & 2.69 & 3.58 & 0.322 $\pm$ 0.004\\
 (14.83, 4.27)& 7.55 - 7.72 & 7.00 - 7.14 & 6.91 & 6.39 & 6.76 & 1.301 $\pm$ 0.014\\
 (38.11, 14.95)& 2.65 - 2.69 & 2.48 - 2.52 & 2.42 & 2.27 & 2.40 & 0.163 $\pm$ 0.002 \\
 (45.63, 45.63)& 1.47 - 1.50 & 1.39 - 1.41 & 1.35 & 1.33 & 1.33 & 0.046 $\pm$ 0.001\\
\hline\hline
\end{tabular}
\end{table*}

\begin{figure*}
	\centering
	\includegraphics[width=7.5cm]{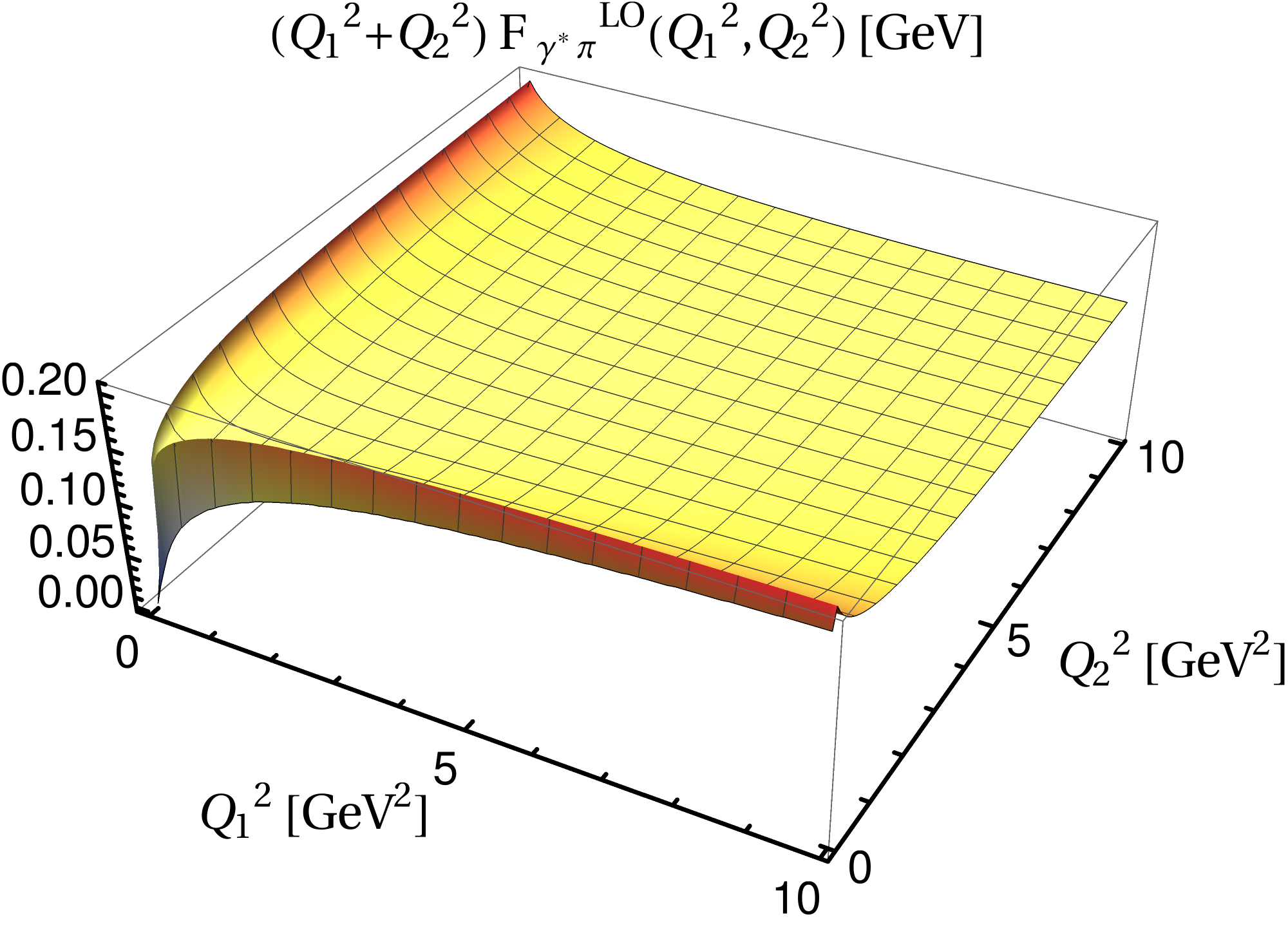}
	\hspace{0.1cm}
	\includegraphics[width=7.5cm]{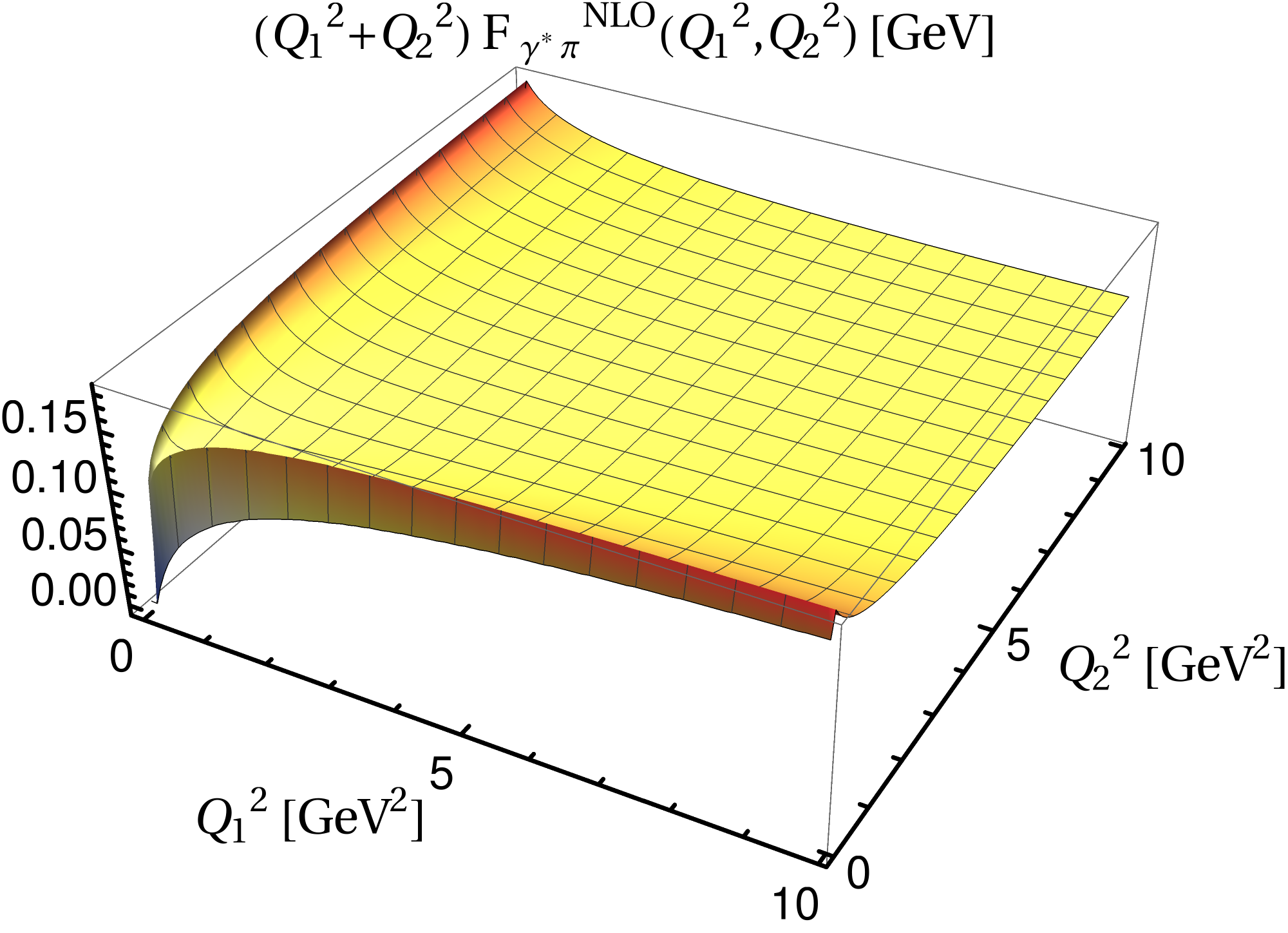}
	\caption{The three-dimensional plots for the $\pi^0\to \gamma^*\gamma^*$ transition form factor obtained from Eq.~(\ref{eq:DTFF_convo}) with BLFQ basis truncation $N_{\rm max}=8$ and $L_{\rm max}=32$. Left and right panels represent the results with LO and NLO hard scattering amplitude, respectively. Note the difference in the scales on the vertical axes. The effect of the scale evolution in the DA is considered following Eq.~(\ref{eq:evolDA}).}
	\label{fig:DTFF2}
\end{figure*}

The numerical results for $F_{\pi \gamma^*}(Q_1^2,Q_2^2)$ within the BLFQ-NJL model calculated at LO and NLO for selected $(Q_1^2,Q_2^2)$ values are given in Table~\ref{tab:DTFF}. We compare our results with results from the light-front quark model (LFQM) and with results from LO and NLO pQCD~\cite{Choi:2019wqx}. We find that our predictions are close to these results as summarized in Table~\ref{tab:DTFF}. On the other hand, our predictions are very different from the VMD model~\cite{BaBar:2018zpn,Choi:2019wqx}:
\begin{align}
F^{\rm VMD}_{{\rm \pi}\gamma^*}(Q^2_1,Q^2_2) 
= \frac{ F_{{\rm \pi}\gamma}(0,0)}{(1+ Q^2_1/\Lambda^2_{\rm \rho})(1+ Q^2_2/\Lambda^2_{\rm \rho})}\,,
\end{align}
where $\Lambda_{\rm \rho}=775$ MeV corresponding to the $\rho$-pole, and $F_{\pi\gamma}(0,0)=0.272(3)$ GeV$^{-1}$~\cite{ParticleDataGroup:2018ovx}. The different behavior between our BLFQ-NJL model result and the VMD model prediction can be attributed to the fact that the TFF in the BLFQ-NJL model exhibits $F_{\pi\gamma^*}(Q^2_1, Q^2_2)\sim 1/(Q^2_1 + Q^2_2)$ when $(Q^2_1, Q^2_2)\to\infty$, which is consistent with the pQCD prediction~\cite{Lepage:1980fj,Braaten:1982yp}, while the TFF in the VMD model behaves as $F^{\rm VMD}_{\pi\gamma^*}(Q^2_1, Q^2_2)\sim 1/(Q^2_1 Q^2_2)$. The consistency between our BLFQ-NJL result and pQCD  is expected as the factorization formula in Eq.~(\ref{eq:DTFF_convo}) is applied in our method. Interestingly, for the
singly virtual TFF, i.e, for $Q^2_1=0$ or $Q^2_2=0$, both the models show the expected scaling behavior $F_{\pi\gamma^*}(Q^2,0)\sim 1/Q^2$ at the large $Q^2$ regime. 

We show the three-dimensional plots for $(Q^2_1 + Q^2_2) F_{\pi\gamma^*}(Q^2_1,Q^2_2)$ calculated at LO and NLO in Fig.~\ref{fig:DTFF2}. The qualitative behavior is found to be consistent with the LFQM result~~\cite{Choi:2019wqx}. As can be seen from the figure, our BLFQ-NJL model results also show the same scaling behavior as predicted by pQCD~\cite{Lepage:1980fj,Braaten:1982yp}. 

\section{Summary}
We have evaluated the valence-quark distribution amplitude from the light-front wave functions of the pion in the framework of the basis light front quantization. Our result is based on the wave functions as the eigenfunctions of an effective Hamiltonian which includes the confinement potentials and the color-singlet Nambu–Jona-Lasinio interactions. The meson DA then evolves according to the ERBL evolution equation from pQCD. We have analyzed the QCD evolution of our pion DA and found that 
it agrees well with the FNAL-E-791 data. At the scale of 10 GeV$^2$, our DA is close to the asymptotic QCD prediction. 

The non-vanishing moments of the DA in BLFQ framework have been found to be consistent with various theoretical predictions. We have also investigated the sensitivity of the pion DA, decay constants, and the radiative decay width to the BLFQ basis size and found good convergence towards the experimental data with increasing longitudinal basis size.

We have calculated the singly and doubly virtual pion-photon TFFs  for $\pi^0\rightarrow \gamma^*\gamma$ and $\pi^0\rightarrow \gamma^*\gamma^*$ transitions using the DA with the hard-scattering formalism. We have taken into account the LO evolution effects of the pion DA and NLO corrections to the hard scattering amplitude for the calculations at finite $Q^2$. We have  found that our prediction for $Q^2F_{\pi\gamma}(Q^2)$ agrees reasonably well with the Belle Collaboration data, but it disagrees  with the rapid growth of the large $Q^2$ pion-photon TFF data reported by the BaBar Collaboration. In the meantime, we have observed that the doubly virtual pion-photon TFF in the BLFQ-NJL model is consistent with the LFQM and the pQCD predictions. Our BLFQ-NJL model result for the doubly virtual TFF manifests as
$F_{{\rm \pi}\gamma^*}(Q^2_1, Q^2_2)\sim 1/(Q^2_1 + Q^2_2)$ when $(Q^2_1, Q^2_2)\to\infty$, which agrees with the pQCD prediction.

\section*{Acknowledgements}
C. M. is supported by new faculty start up funding by the Institute of Modern Physics, Chinese Academy of Sciences, Grant No. E129952YR0. 
C. M. and S. N. thank the Chinese Academy of Sciences Presidents International Fellowship Initiative for the support via Grants No. 2021PM0023 and 2021PM0021, respectively.  X. Z. is supported by new faculty startup funding by the Institute of Modern Physics, Chinese Academy of Sciences, by Key Research Program of Frontier Sciences, Chinese Academy of Sciences, Grant No. ZDB-SLY-7020, by the Natural Science Foundation of Gansu Province, China, Grant No. 20JR10RA067 and by the Strategic Priority Research Program of the Chinese Academy of Sciences, Grant No. XDB34000000.
 S. J. is supported by U.S. Department of Energy, Office of Science, Office of Nuclear Physics, contract no. DE-AC02-06CH11357. J. P. V. is supported in part by the Department of Energy under Grants No. DE-FG02-87ER40371, and No. DE-SC0018223 (SciDAC4/NUCLEI). 

\end{document}